\documentclass[namedreferences]{solarphysics}

\usepackage[optionalrh]{spr-sola-addons}
\usepackage{graphicx}                    
\usepackage{amssymb}                    
\usepackage{color}                       
\usepackage{amsbsy}

\usepackage[english]{babel}

\newcommand{\adv}      {{\it Adv. Space Res.}}

\newcommand{\aap}      {{\it Astron. Astrophys.}}

\newcommand{\aj}       {{\it Astron. J.}}
\newcommand{\apj}      {{\it Astrophys. J.}}

\newcommand{\solphys}  {{\it Solar Phys.}}

\begin{document}

\begin{article}

\begin{opening}
\title{Examination of artifact in vector magnetic field SDO/HMI measurements}

\author[addressref=aff1,corref,email={rud@iszf.irk.ru}]{\inits{G.V.}\fnm{G. V.}~\lnm{Rudenko}}

\author[addressref=aff1,corref,email={dmitrien@iszf.irk.ru}]{\inits{I.S.}\fnm{I.S.}~\lnm{Dmitrienko}}
\address[id=aff1]{Institute of Solar-Terrestrial Physics SB RAS, Lermontov St. 126,
Irkutsk 664033, Russia}

\runningauthor{G.V. Rudenko and I.S. Dmitrienko}
\runningtitle{Examination of artifact in vector magnetic measurements}

\begin{abstract}

In this paper, we came to conclusion that there is a significant systematic error in the SDO/HMI vector magnetic data, which reveals itself in a significant deviation of the lines of the knot magnetic fields from the radial direction. The value of this deviation demonstrates a clear dependence on the distance to the disk center. This paper suggests a method for correction of the vector magnetograms that eliminates the detected systematic error.

\end{abstract}
\end{opening}

\section{Introduction}
     \label{S1}

Satellite vector magnetic data from the SDO/HMI spacecraft
represent a significant breakthrough in solar magnetography.
Spatial resolution, quality of full disk vector magnetograms,
regularity and high duty cadence of observations have no analogue
neither in Earth, nor in satellite measurements. Invaluable is the
contribution that can be made in the nearest future by the
ever-growing time series of continuous observations for space
weather predictions and fundamental research of magnetic nature of
the solar activity. In particular, we can expect a considerable
improvement in reliability of prediction of the solar wind
parameters and IMF polarity of circumterrestrial space owing to the
possibility of using the new vector synoptic maps \citep{Gosain}.  To reconstruct the current global 3D structure of the
filed in potential approximation and further modeling  of the solar
wind, longitudinal field synoptic maps were used in the past
\citep{Harvey}. The advantage of new vector synoptic maps  is
brought about by two aspects. First, employing vector measurements
enables us to extrapolate the field for the boundary field radial
component (Neumann problem). Physically, such definition of the
extrapolation problem is better substantiated comparing with
extrapolation for longitudinal component, because the actual
measurements are taken at the level where both potential
approximation and even more general force-free approximation are
sure not applicable . In these circumstances, different boundary
value problems must inevitably lead to different results of
extrapolation; these results, among the rest will give different
components of a radial field at the boundary. Second, we construct
Br synoptic map as a map for scalar value (unlike $B_{LOS}$ maps). This
is an important point, since reconstruction of synoptic map for a
non-scalar value is not quite correct.

For any large project that puts some new physical data into common
use, it is strategically important remove, if possible, any
significant artificial or natural errors, should such be present
and can be identified. This is desirable to be done either before
employing this information, or in the very beginning. Our work
deals with exactly such kind of problem for the new SDO/HMI data.
Herein we ascertain the fact of existence of a significant
systematic error in the data submitted. We identify clearly this
error from the analysis of the measurements data of knot magnetic
fields that concentrate in the convection cell grid nodes of the
quiet Sun. The observed magnetic knots result from the surrounding
plasma raking up magnetic pipes by horizontal motions; this leads
to magnetic flux concentration and subsequent  radialization of the
field. The compactness property of knot fields and significant
excess of their magnitude relative to background values ($> 500 \ G$)
enables us to select them using a sufficiently simple algorithm.
The knot field inherent radiality is used as the main criterion for
testing the magnetic field measurement data. We show (section 2)
that the same systematic deviation (up to $20$ degrees) of the knot
field from the radial direction toward the limb is revealed in all
SDO/HMI vector magnetograms. This deviation depends on the distance
to the disk center. Since observation result must not depend on
observer position, we conclude that the revealed dependence can
only have an artificial cause; this cause is likely unrecoverable
in modern technologies for receiving and processing Stokes
parameters used to obtain final values of vector magnetic field.
In Section 3 we propose the idea of correcting the initial
original vector magnetograms, based on the assumption that the
systematic deviation from the radiality should be absent, and its
presence is consequence of the presence of an error in the data of
the angle of the field  inclination   relative to the line of
sight. The manifestations of this error in the knots fields lead to
the observed dependence of the deviation from the radiality on the
position of the measurement point on the disk.
We show that our correction almost eliminates effects of unnatural
behavior of knot fields. Unlike the original magnetograms, the
corrected ones do not contradict the results of the virial theorem
for a nonlinear field \citep{Livshits}(Livshits et al.2015)-"virial" energy is
positive (first of all), it exceeds the energy of the reference
potential field.

\section{Identification and Analysis of Knot Fields }
     \label{S2}
\subsection{Selection and Geometric Interpretation of Knot Magnetic Fields}\label{S2.1}
Our examination relies on the natural assumption about radiality of isolated small-scale magnetic structures of large magnitudes. Most of such structures correspond to the knots that concentrate in the convection cell grid nodes of the quiet Sun.   Selection of the structures of our interest and determination of  their physical properties from the magnetograms can be described with the Selection Algorithm (SA) as follows: \\
-- using the IDL procedure "LABEL\_REGION",  we choose the full set of isolated regions with the $\left|{\bf B}\right|>300 \ G$ values; \\
-- from the obtained set, we select only the $A_n$ regions with pixel number not more than $35$ and $max\left|{\bf B}(A_n)\right|>500 \ G$;\\
-- for each $A_n$ we find pixel $\left(i^n_{max},j^n_{max}\right): \left|{\bf B}\left(i^n_{max},j^n_{max}\right)\right|=max|{\bf
B}(A_n)|$; \\
-- to each $A_n$ we assign value of magnetic field ${\bf B}^n = {\bf B} (i_{max}^n, j_{max}^n)$ and radius-vector $r^n = r (i_{max}^n, j_{max}^n)$.\\
The values of $B^n$ and $r^n$ are used for further analysis. For simplicity, we will further omit index.

\begin{figure}
\centerline{\includegraphics[width=1.\textwidth,clip=]{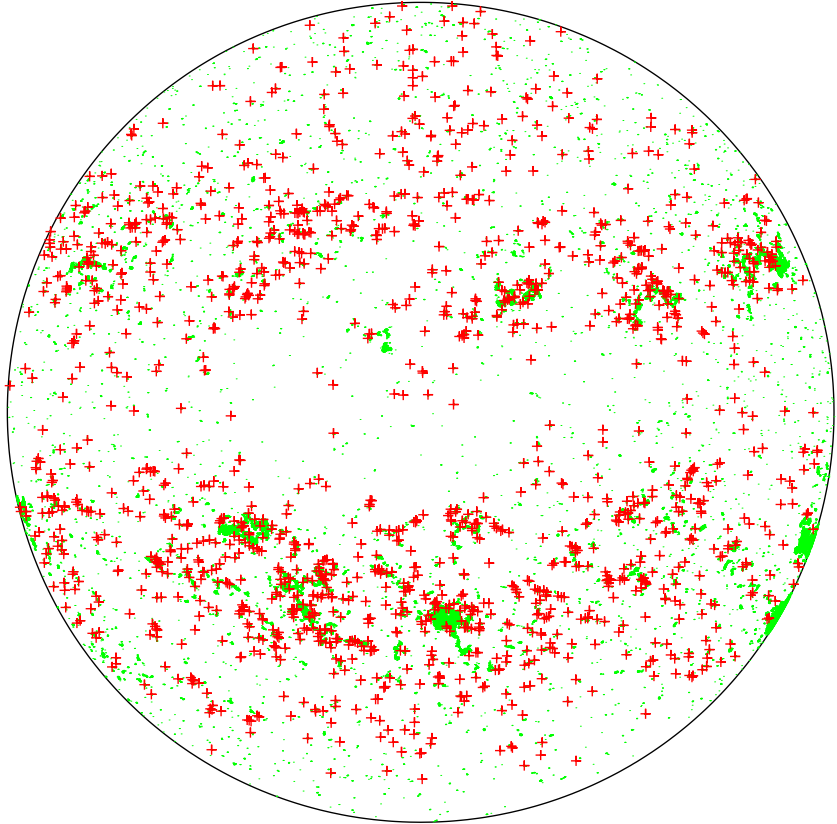}}
\caption{SDO/HMI 2012-01-1505:12:00UT. Typical distribution of magnetic knots over the solar disk: $|{\bf B}|>300  \ G$ –green points, knots – red crosses.} \label{kfig1}
\end{figure}

\begin{figure}
 \centerline{\includegraphics[width=0.7\textwidth]{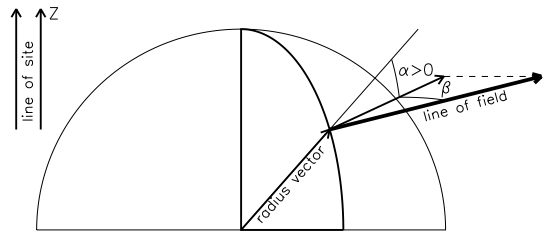}}
 \caption{Angular characteristics of the deviation from the magnetic field vector radiality.}
 \label{kfig2}
 \end{figure}

Figure \ref{kfig1} shows a typical example of defining location of the centers of knot regions (highlighted with red crosses). The detected knot regions are distributed evenly enough throughout the entire regions of the quiet Sun; they have slight higher concentration near active regions (green).
To assess the degree of radiality of the field of the knot regions, we will use the following three values (knot parameters $B_r^{LOS}/|{\bf B}| ,  \alpha, \beta)$: \\
-- $B_r^{LOS}/|{\bf B}|$ is
\begin{equation}\label{keq1}
B_r^{LOS}/|{\bf B}|=\frac{B_{LOS}}{\cos\mu}/|{\bf B}|,
\end{equation}
where $B_{LOS}=B_z$ is projection of vector {\bf B} along the line of sight, $\mu$ is the angle between the line of sight and the radius-vector of the knot location on the disk ($B_r^{LOS}=|B_r |=|{\bf B}|$ when the field is exactly radial); \\
-- $\alpha$     is the angle between the observation point radius-vector and magnetic field component in meridional plane that is determined by the observation point radius-vector and the line of sight $z$-axis ($\alpha > 0$ toward the limb, $\alpha=0$ when the field is exactly radial); \\
-- $\beta$ is the angle between the vector of field line (always believed to be directed away from the Sun) and meridional plane ($\beta=0$ when the field is exactly radial).
Figure \ref{kfig2} shows geometry of the angles $\alpha$ and $\beta$.

 It is important to note that to derive the value $B_r^{LOS}/|{\bf B}|$  we don't need the information about the azimuthal field direction. Only, we need to eliminate $\pi$-uncertainty of the  $\phi$ azimuth to obtain the angular knot parameters $\alpha$ and $\beta$ . In this paper, we used the method described in \citep{ambig}  to eliminate the $\pi$-uncertainty.
\subsection{Distribution of HMI Knot Parameters Over the Disk}\label{S2.2}
\begin{figure}
\centerline{\includegraphics[width=0.7\textwidth]{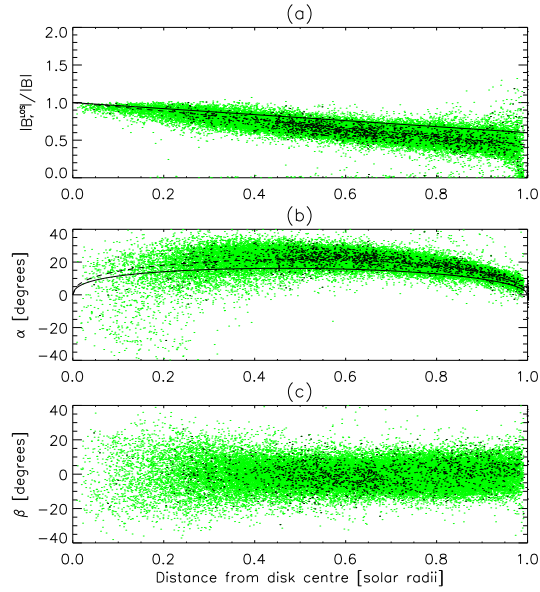}}
\caption{Dependence of knots parameters on the distance to the solar disk center: (a) - $|B_r^{LOS}|/|{\bf B}|$; (b) - $\alpha$, (c) - $\beta$;
green - for the full set of magnetograms, black - points of one magnetogram SDO / HMI 2012-01-15 05:12:00 UT.}
\label{kfig3}
\end{figure}

In our paper we examined the knot parameters characterizing deviation of the knot field from radiality for quite a large set of randomly time-selected SDO/HMI magnetograms ($= 30$) for 2012-2014. In all magnetograms, parameters $B_r^{LOS}/|{\bf B}|$, $\alpha$ and $\beta$ demonstrate factually identical behavior on the solar disk (see Figure \ref{kfig1})
From Figure \ref{kfig3}, it follows that only the $\beta$ parameter demonstrates the expected – in case there are no data artifacts – statistic behavior with zero-mean.
The other parameters $|B_r^{LOS}|/|{\bf B}|$, $\alpha$ show a clear dependence of their averages on the distance to the solar disk center. On the one hand, non-zero $|B_r^{LOS}|/|{\bf B}|$,  $\alpha$  contradict the hypothesis of radiality of  the knots fields; on the other hand, their dependence on the distance to the center of the solar disk indicates the presence of an error of artificial origin in the data. Indeed, even if we exclude the property of radiality in the selected magnetic elements, in this case there should be, in principle, no statistical connection between knots parameters and the distance to the center of the solar disk. The same magnetic elements can not give different values of the field, depending on their visible position on the solar disk. It is also clear that the error we reveal can not be a consequence of the subsequent processing of the initial magnetograms associated with the method of solving the π-uncertainty problem of the transverse magnetic component of the measurements. In the case of an error generated only by solving the π-uncertainty problem, we would not observe any    relationship with the distance of $|B_r^{LOS}|/|{\bf B}|$, which does not have $\pi$-uncertainty. Note   the maximum deviation of the angle α is of the order of $20$ degrees at medium distances and the decrease of the $|B_r^{LOS}|/|{\bf B}|$ - ratio  near the limb almost is $2$ times; that  indicate the essentiality of the artificial error magnitudes.

Let us take notice of the apparent simplicity of relationship between the $|B_r^{LOS}|/|{\bf B}|$  value and the distance; in Figure \ref{kfig3} (a), this relationship is visually perceived as close to linear. Let us write it down in the form
\begin{equation}\label{keq2}
B_r^{LOS}/|{\bf B}|=1-kr_\perp,
\end{equation}
where $r_\perp$ is the distance to the disk center in solar radii. On the full set of magnetograms, we obtained a fitting  value of the Formula (\ref{keq2}) linear coefficient, this fitting value equals
\begin{equation}\label{keq3}
k=k_{fit}=0.565,
\end{equation}
Figure \ref{kfig3} (a) shows the relevant dependence as a dashed line (we will comment later on the solid line in Figure \ref{kfig3} (a) and the  lines in Figure \ref{kfig3} (b)).

Note that our results, shown in Figure \ref{kfig3} (as well as (\ref{keq2}) with (\ref{keq3}) coefficient), explain well the Figure 4b from \cite{Leka}, which demonstrates the $|B_r^{LOS}|$ dependence on magnitude $|B_r|$ in the polar region. Statistical distribution of these values clearly shows approximately two-fold decrease of the first value relating to the second one (same as in our case for $r_\perp \sim 1 $(Figure \ref{kfig3} (a))

Thus, the presence of a significant systematic error in the HMI/SDO data can be deemed a proved fact.
\section{Correction of "Knot" Systematic Error}
\label{S3}
A very simple dependence of the knot parameter $|B_r^{LOS}|/|{\bf B}|$ on the distance $r_\perp$ (Figure \ref{kfig3} (a)) gives reason to think about the possibility of finding some correction method that will, at least formally, eliminate the discovered statistical effects. Initially, two approaches to resolve this problem can be suggested: "geometric" – correction depends on the measured element location on the disk; "local" – correction depends only on the measured value themselves and does not depend on the measured element location on the disk. The first approach has proved to be quite problematic. At least, we could not find any reasonable option to implement it. We propose a correction based on the second assumption, it is quite simple, and preliminary yields quite reasonable results. We hope that the new magnetograms, which unlike the original ones have no revealed deficiencies, are more suitable in their future practical use: to obtain both the corona global model \citep{Svalgaard, Wang, Riley1, Riley2}, and nonlinear simulation of active magnetic regions \citep{Sun, Thalmann, Tadesse}(Sunetal. 2012; Thalmann et al. 2012; Tadesse et al. 2013).
\subsection{Method}
\label{S3.1}
We introduce the following notations: \\
-- $|{\bf B}^*|$ is modulus of true (required) magnetic field; \\
-- $\gamma^*$ is inclination of true magnetic field;\\
-- $\phi^*$ is azimuth of transverse true magnetic field;\\
-- $B_{LOS}^*=|{\bf B}^*|\cos\gamma^* $ is longitudinal component of true magnetic field
Relevant notations without asterisks will be referred to first measured field parameters. We believe the following identity relations valid for measurements in every point of the disk:
\begin{equation}\label{keq4}
\phi^* \equiv \phi,
\end{equation}
\begin{equation}\label{keq5}
B_{LOS}^*=|{\bf B}|^*\cos\gamma^* \equiv B_{LOS}=|{\bf B}|\cos\gamma.
\end{equation}
Suppose, the true field is radial. In that case, we have for the knots in Formula (\ref{keq2})
\begin{equation} \label{keq6}
r_\perp=\sin\gamma^*,
\end{equation}
\begin{equation} \label{keq7}
\cos\mu=\cos\gamma^*.
\end{equation}
Using (\ref{keq1}), (\ref{keq6}), (\ref{keq7}), we can rewrite formula (\ref{keq2}) ) as follows:
 \begin{equation} \label{keq8}
|B_r^{LOS}| / |{\bf B }|=\frac{|{\bf B }|\cos\gamma}{\cos\gamma^*|{\bf B }|}=\frac{\cos\gamma}{\cos\gamma^*}=1-k\sin\gamma^*.
\end{equation}
Equation (\ref{keq8}) provides a relationship between the "true" and measured inclination. The $\gamma^*$ value can be found using the Newton algorithm. After $\gamma^*$ is found, we derive the "true" field modulus from (\ref{keq5}) and (\ref{keq7}):
\begin{equation} \label{keq9}
|{\bf B }^*|=|{\bf B }|\frac{\cos\gamma}{\cos\gamma^*}=|{\bf B }|(1-k\sin\gamma^*).
\end{equation}
Formulas (\ref{keq4}), (\ref{keq8}), (\ref{keq9}) allow us to uniquely determine new values of the field modulus, inclination and azimuth ($|{\bf B }^*|$, $\gamma^*$, and $\phi^*$) basing on their original values and the $k$ value. Extending these relationships to all the points of the solar disk, we obtain a magnetogram with new inclination and field modulus, while the longitudinal field component and transverse field azimuth remain the same. Therewith, the average of the $|B_r^{*LOS}| / |{\bf B }^*|$  value in knot regions shall accept value 1 everywhere. It is important to note that the correction based on formulas (\ref{keq4}), (\ref{keq8}), (\ref{keq9}) does not depend on procedure of $\pi$-disambiguation of the azimuth $\phi$ of transverse field; this correction can be done before this procedure.
\subsection{Distribution of Knot Parameters for Corrected Magnetogram}
\label{S3.2}
\begin{figure}
\centerline{\includegraphics[width=0.7\textwidth]{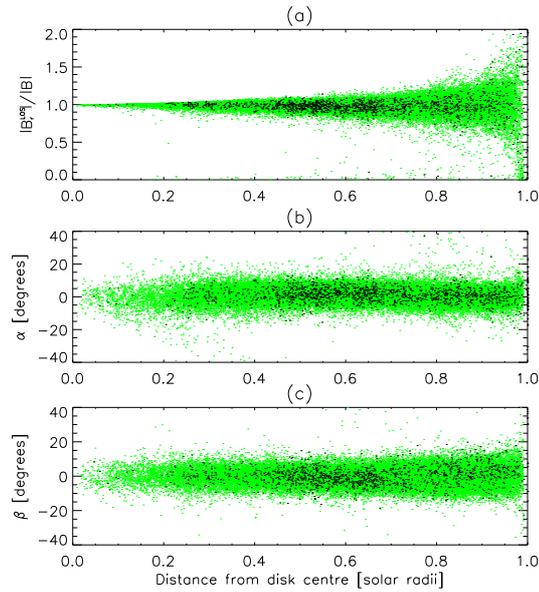}}
\caption{The corrected dependence of knot parameters on the distance to the solar disk center ($k=0.565$):(a) - $|B_r^{LOS}|/|{\bf B}|$ ; (b) - $\alpha$, (c) - $\beta$;
 green - for the full set of magnetograms, black - points of one magnetogram SDO/HMI 2012-01-15 05:12:00 UT. SA (non-corrected magnetogram}
\label{kfig4}
\end{figure}

First, we consider the correction result shown in Figure \ref{kfig4}, with the k value from (\ref{keq3}), for knot regions preliminary highlighted in the non-corrected magnetograms. As expected, the dependence between the knot parameter $|B_r^{LOS}| / |{\bf B }|$  and the distance is eliminated sufficiently well.
We observe that only this value dispersion depends on the distance to the solar disk center. In addition, the second knot parameter $\alpha$ dependence on the distance to the center of the solar disk disappears, too. In fact, pursuant to the $\alpha$ definition, from (\ref{keq2}) we can predict its behavior on the disk (before correction), believing $\beta=0$:
\begin{equation} \label{keq10}
\alpha=\arccos\left((1-kr_\perp)(1-r^2_\perp)-r_\perp\sqrt{1-(1-kr_\perp)^2(1-r^2_\perp)}\right).
\end{equation}
Figure \ref{kfig3} (b) depicts dependence of (\ref{keq10}) with a dashed line for $k=0.565$. It is clear that with the disappearance of (\ref{keq2}) dependence, the average $\alpha$ must be zero everywhere.

\begin{figure}
\centerline{\includegraphics[width=0.7\textwidth]{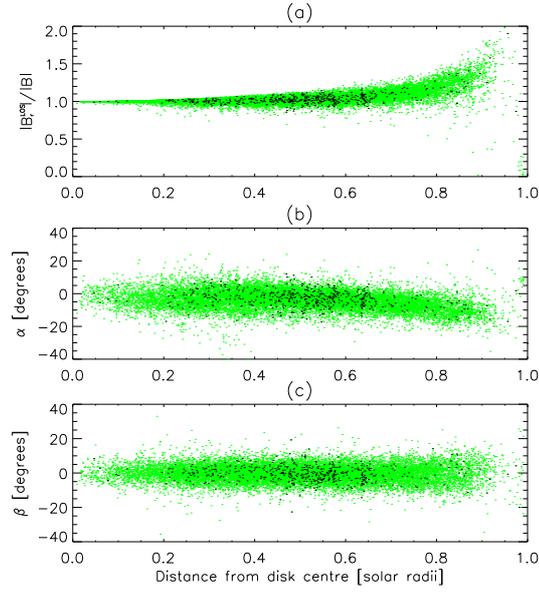}}
\caption{The corrected dependence of knot parameters on the distance to the solar disk center ($k=0.565$):(a) - $|B_r^{LOS}|/|{\bf B}|$ ; (b) - $\alpha$, (c) - $\beta$;
green - for the full set of magnetograms, black - points of one magnetogram SDO / HMI 2012-01-15 05:12:00 UT. SA (corrected magnetogram)}
\label{kfig5}
\end{figure}

\begin{figure}
\centerline{\includegraphics[width=0.7\textwidth]{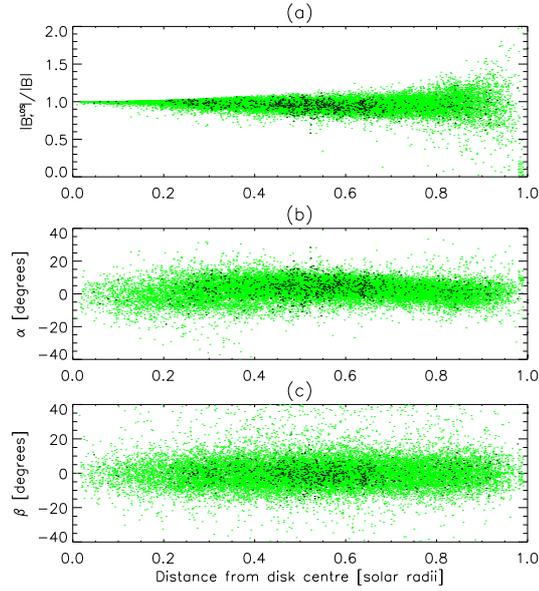}}
\caption{The corrected dependence of  knot parameters on the distance to the solar disk center ($k=0.4$):(a) - $|B_r^{LOS}|/|{\bf B}|$ ; (b) - $\alpha$, (c) - $\beta$;
green - for the full set of magnetograms, black - points of one magnetogram SDO / HMI 2012-01-15 05:12:00 UT. SA (corrected magnetogram)}
\label{kfig6}
\end{figure}

Despite the fact that the correction with the selected value $k$ led to an improvement in the behavior of the knots parameters on the disk (Figure \ref{kfig4}), it is not completely correct.
For selection of the knot regions in SA we used the non-corrected ("wrong") values of the field modulus, which changes significantly  with the correction as we determined. Hence, the knot parameters obtained from the corrected magnetogram for $k=0.565$ give the results with some residual statistical dependence on the distance, near the limb (see Figure \ref{kfig5}).
This result is quite natural, because the obtained fitting-valuek was derived from the set of "wrong magnetograms". As a result of selection we found $k=0.4$ as the most appropriate value (solid line in Figure. \ref{kfig3} (a)). As Figure \ref{kfig6} shows, correction with the derived $k$ value factually takes off the knot parameters dependence on the distance to the solar disk center. Figure \ref{kfig3} (b) shows dependence (\ref{keq10}) corresponding to this value $k$ as a solid line. It is natural that the latter is a bit offset from the $\alpha$ knot parameter averages.

\begin{figure}
\centerline{\includegraphics[width=0.7\textwidth]{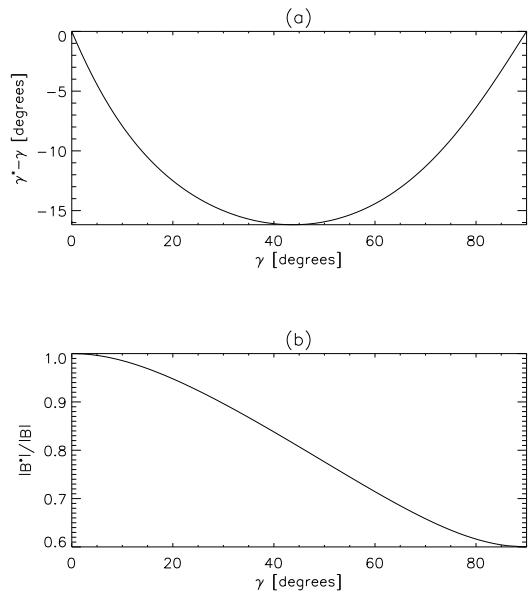}}
\caption{Plots that depict variations in inclination (a) and field modulus (b) after correction}
\label{kfig7}
\end{figure}

So, we have magnetograms whose knot fields satisfy the natural assumption of radiality. In any point of magnetogram, changes in modulus and inclination depend only on the $\gamma$ local value, according to formulas (\ref{keq8}), (\ref{keq9}) for $k=0.4$. These variations change inclinations and magnetic field significantly (see Figure \ref{kfig7}).
Only results of application solutions of physical problems can later allow us to assess how legitimate such correction is for magnetic parameters in active regions, because in a general way, we do not have any assumptions about true orientation of the magnetic field.
\subsection{Correspondence of Magnetograms With the Force-free Approximation}
\label{S3.3}
Some feature of the proposed correction can be given on the basis of the assumed force-free nature of magnetic field. At photospheric heights, this assumption must be approximately fulfilled at least for the regions with strong magnetic field. In force-free approximation, the virial theorem is valid, it gives the full energy equation as the surface integral for the entire Sun sphere (see  \cite{Livshits}):
\begin{equation} \label{keq11}
E\equiv\frac{1}{4\pi}\int_V|{\bf
B}|^2dv=E_{vir}=\frac{R_{sun}}{8\pi}\int_S(B^2_r-B^2_t)ds,
\end{equation}
where $B_t$ is a tangential component of magnetic field.

As discussed in paper by \cite{Livshits}, minimal energy of reference potential field ($B_r^{pot} = B_r$) means that any non-linear field has a greater radiality relative to its reference field:
\begin{equation} \label{keq12}
\int_SB^2_tds<\int_S(B^{pot}_t)^2ds,
\end{equation}
Calculating the integrals (\ref{keq12}), we can check to what extent the original and corrected magnetograms satisfy condition (\ref{keq12}), and energy positivity condition (\ref{keq11}) (we shall call $E_{vir}$ virial energy).
It should be noted that the appropriate testing is quite conditional, for the following reasons:\\
-- restriction due to integration only on the visible part of the sphere;\\
-- high noise of transverse field measurements;\\
-- errors related to principle impossibility to solve the problem of transverse field $\pi$-uncertainty in weak fields of quiet Sun regions due to high noise;\\
-- mainly a non-force-free nature of the field in weak field regions.

\begin{figure}
\centerline{\includegraphics[width=0.7\textwidth]{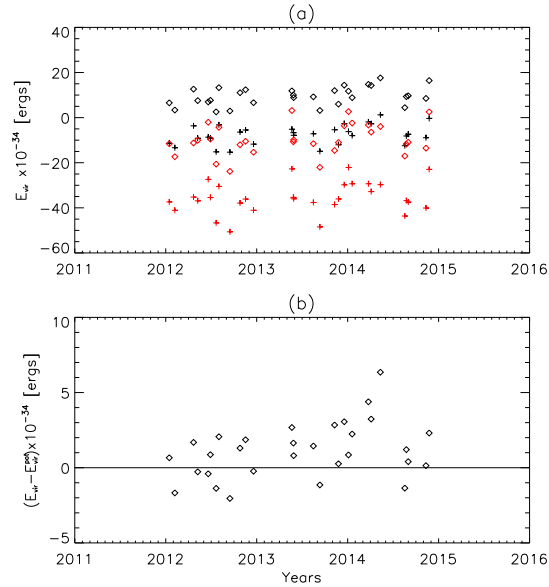}}
\caption{(a) - calculated magnitudes of virial energies for the full set of magnetograms,$E_{vir}$-  crosses,$E_{vir}(|{\bf(B)}|> 150$  $erg)$ -  rombs, red - non-corrected magnetograms, black - corrected magnetograms ($k = 0.4$); (b) - "free
energy".}
\label{kfig8}
\end{figure}

Despite the listed restrictions of such testing, the results presented in Figure \ref{kfig8}, in our view, still prove specific positive features of magnetogram correction that we offer.
We see that original magnetograms mainly provide "wrong" negative magnitudes of virial energy (both for full magnetograms and these with  cutting condition $|{\bf(B)}|> 150$ $erg)$    to eliminate the noise). In case the noise is cut, the magnetograms after correction demonstrate "valid" positive values of virial energy, in most cases these values exceed potential energy. It is interesting to note that the  cases of "free energy" negative values show magnetograms with "cut" active regions – ascendant or descendant within the visibility region.
\section{Conclusion}
\label{S4}

We have shown that the vector magnetic data from Helioseysmic magnetic imager on-board Solar dynamic observatory (SDO/HMI) contain significant systematic error. It becomes apparent in the fact that the magnetic field in small-scale magnetic elements with high field intensity (magnetic knots) deviates from the radial direction toward the solar limb. The deviation value depends on the distance to the center of the visible solar disk and reaches maximum of $\sim 20$ degrees at distances of about $0.4$ of the solar radius to the disk center.

We offer the correction that eliminates the revealed systematic error. This correction is preliminary and requires further approbation on specific application problems. Perhaps it can serve to find causes of the data systematic error and to eliminate this error at the hardware level.
\

\begin{acks}
 This study was supported by the Russian Foundation of Basic Research under grants 15-02-01077;
was supported by the Program of basic research of the RAS
Presidium No. 7.
\end{acks}

\begin{acks}[Disclosure of Potential Conflicts of Interest]
 The authors declare that they have no conflicts of interest.
\end{acks}


\begin{thebibliography}{}
\bibitem[\protect\citeauthoryear{Gosain et al.}{2013}]{Gosain}
Gosain,~S., Pevtsov,~A.~A., Rudenko,~G.~V., Anfinogentov,~S.~A.:
2013, {\apj} \ \textbf{772}, 52. doi:10.1088/0004-637X/772/1/52

\bibitem[\protect\citeauthoryear{Harvey et al.}{1980}]{Harvey}
Harvey,~J., Gillespie,~B., Miedaner,~P., Slaughter,~C.: 1980, {\it
NASA STI/Recon Technical Report} \ No., 81, 21003

\bibitem[\protect\citeauthoryear{Leka et al.}{2017}]{Leka}
Leka,~K.~D., Barnes,~G., Wagner,~E.~L.: 2017, {\solphys} \
\textbf{292}, 36.

\bibitem[\protect\citeauthoryear{Livshits et al.}{2015}]{Livshits}
Livshits,~M.~A., Rudenko,~G.~V., Katsova,~M.~M., Myshyakov,~I.~I.:
2017, {\adv} \ \textbf{55}, 920. doi:10.1016/j.asr.2014.08.026

\bibitem[\protect\citeauthoryear{Riley et al.}{2006}]{Riley1}
Riley,~P., Linker,~J.~A., Miki'c,~Z., Lionello,~R.,
Ledvina,~S.~A., Luhmann,~J.~G.: 2006, {\apj} \ \textbf{653}, 1510.
doi:1510. doi:10.1086/508565

\bibitem[\protect\citeauthoryear{Riley et al.}{2014}]{Riley2}
Riley,~P., Ben-Nun,~M., Linker,~J.~A., Miki'c,~Z., Svalgaard,~L.,
Harvey,~J., Bertello,~L., Hoeksema,~T., Liu,~Y., Ulrich,~R.: 2014,
{\solphys} \ \textbf{289}, 769. doi:10.1007/s11207-013-0353-1

\bibitem[\protect\citeauthoryear{Rudenko and Anfinogentov}{2014}]{Rudenko}
Rudenko,~G.~V., Anfinogentov,~S.~A.: 2014, {\solphys} \
\textbf{289}, 1499. doi:10.1007/s11207-013-0437-y

\bibitem[\protect\citeauthoryear{Sun et al.}{2012}]{Sun}
Sun,~X., Hoeksema,~J.~T., Liu,~Y., Wiegelmann,~T., Hayashi,~K.,
Chen,~Q., Thalmann,~J.: 2012, {\apj} \ \textbf{748}, 77.

\bibitem[\protect\citeauthoryear{Svalgaard et al.}{1978}]{Svalgaard}
Svalgaard,~L., Duvall,~T.~L.~Jr., Scherrer,~P.~H.: 1978,
{\solphys} \ \textbf{58}, 225. doi:10.1007/BF00157268

\bibitem[\protect\citeauthoryear{Tadesse et al.}{2013}]{Tadesse}
Tadesse,~T., Wiegelmann,~T., Inhester,~B., et al: 2013, {\aap} \
\textbf{550}, A14.

\bibitem[\protect\citeauthoryear{Thalmann et al.}{2012}]{Thalmann}
Thalmann,~J.~K., Pietarila,~A., Sun,~X., Wiegelmann,~T.: 2012,
{\aj} \ \textbf{144}, 33.


\bibitem[\protect\citeauthoryear{Wang and Sheeley}{1992}]{Wang}
Wang,~Y., Sheeley,~J.~N.~R.: 1992, {\apj} \ \textbf{392}, 310.
doi:10.1086/171430


\end{thebibliography}

\end{article}

\end{document}